\documentclass[aps,prl,twocolumn,groupedaddress,showpacs]{revtex4}

\usepackage{graphicx}
\usepackage{amssymb}
\usepackage{epstopdf}

\usepackage{amssymb}

\usepackage{graphicx}

\usepackage{color}

\newcommand{\comment}[1]{}

\newcommand{\micron}{\mu\text{m}}
\newcommand{\kt}{k_{\text{B}}T}

\begin{document}

\bibliographystyle{apsrev}

% Use the \preprint command to place your local institutional report
% number in the upper righthand corner of the title page in preprint mode.
% Multiple \preprint commands are allowed.
% Use the 'preprintnumbers' class option to override journal defaults
% to display numbers if necessary
%\preprint{}

%Title of paper
\title{Sedimentation of active colloidal suspensions }

% repeat the \author .. \affiliation  etc. as needed
% \email, \thanks, \homepage, \altaffiliation all apply to the current
% author. Explanatory text should go in the []'s, actual e-mail
% address or url should go in the {}'s for \email and \homepage.
% Please use the appropriate macro foreach each type of information

% \affiliation command applies to all authors since the last
% \affiliation command. The \affiliation command should follow the
% other information
% \affiliation can be followed by \email, \homepage, \thanks as well.
\author{J\'{e}r\'{e}mie Palacci}
\author{C\'{e}cile Cottin-Bizonne}
\author{Christophe Ybert}
\author{Lyd\'{e}ric Bocquet}
%\email[]{Your e-mail address}
%\homepage[]{Your web page}
%\thanks{}
%\altaffiliation{}
\affiliation{LPMCN;  Universit\'e de Lyon; Universit\'e Lyon 1 and
CNRS, UMR 5586; F-69622 Villeurbanne, France\\}

%Collaboration name if desired (requires use of superscriptaddress
%option in \documentclass). \noaffiliation is required (may also be
%used with the \author command).
%\collaboration can be followed by \email, \homepage, \thanks as well.
%\collaboration{}
%\noaffiliation

%\date{\today}

\begin{abstract}
In this paper, we investigate experimentally the non-equilibrium steady state of an active colloidal suspension under gravity field.
The active particles are made of chemically powered colloids, showing self propulsion in the presence of an added fuel, here hydrogen peroxide. 
%and
The active suspension is
studied in a dedicated microfluidic device, made of permeable gel microstructures. Both the microdynamics of individual colloids and the global stationary state of the suspension under gravity --  density profiles, number fluctuations --  are measured with optical microscopy. 
%using high speed particle tracking and confocal microscopy.
This allows to connect the sedimentation length to the individual self-propelled dynamics, suggesting that in the present dilute regime the active colloids 
behave as 'hot' particles. Our work is a first step in the experimental exploration of the out-of-equilibrium properties of artificial active systems. 
%This allows to relate the 
%Simultaneously, the stationary density profiles of the suspension is measured using confocal microscopy.
% the sedimentation problem of a dilute active suspension of synthetic microswimmers.  
%It endosses the first steps to revisit the basics of the out of equilibrium statistical physics. 
%{\color{blue} Microdynamics and Stationary sedimentation profiles are measured by optical tracking. Fluctuations}
%%
%We therefore perform a Jean Perrin experiment coupled with a diffusion coefficient measurement as a first step in the exploration of this concept. We probe {\it independantly} the dynamics of the swmimming colloids  with a high speed particle tracking  as well as the response of the system under a force field investigating the colloid density  profile. We measure an enhanced long time diffusion increasing with the colloid swimming velocity and an exponential decay of the  sedimentation with a decay increasing with the swimming velocity. We show a match between these independant measures according to  a schmoluchowski description. This result is the first experimental characterization of the statistical physics ruling an assembly  of artificial self propelled microswimmers.
 \end{abstract}


\maketitle

The collective behavior of 'active fluids', made of self-propelled entities, has raised considerable interest over the recent years in 
the context of non-equilibrium statistical physics \cite{ramaswamy,Vicsek,Joanny,ignacio,Chate,diLeonardo,Marchetti,Tailleur}. Such systems are rather common in living systems, from
%These systems, which encompass systems as diverse as 
swimming cells, bacteria colonies \cite{Lauga,Lauga2, leptos,Libchaber2000,Cisneros}, to flocks of birds or fishes \cite{Vicsek}.
% vibrated granular rods, 
These entities move actively by consuming energy and %thereby energy transduction. 
their behavior is thus intrinsically out of equilibrium. 
Building a general framework describing their collective properties remains accordingly
a challenging task and led
%Their behavior
%raises challenging theoretical questions 
%%about the emergence of large scale order and finding to find 
%in order to  and this led 
to a considerable
amount of work towards this aim \cite{ramaswamy,Vicsek,Joanny,ignacio,Chate,diLeonardo,Marchetti,Tailleur}. %along these lines.
% citer Ramaswamy, Ignacio, Marchetti, Cates
%community. 
%Such systems,
%such as bacteria assemblies, have raised challenging questions due to 
%This has motivated a considerable amount of theoretical work. 
By contrast, much less work has been performed on the experimental side, %much of the recent work has been performed on 
%early studies have explored the behavior of 
and mainly on assemblies of living micro-organisms -- which are naturally self-propelled -- \cite{leptos, Libchaber2000,Cisneros,Rafai}, however at the expense of a lack
of control and flexibility of the individual particles and their interactions. There is therefore a need for new experiments %to develop 
based on
{\it artificial} model systems, 
involving suspensions of microscopic active particles with controled propulsion and interaction mechanisms -- here designated as '{\it active suspensions}'. %self propelled a
Several routes to design individual artificial microscopic swimmers have been explored recently
\cite{Bibette,Paxton,Howse}, taking benefit of the recent
progress made to shape and design colloidal particles at microscales. % ref JANUS + ?
However going from the individual to the many particles situation remains a challenging task and has not been achieved up to now
with artificial motile particles.
%challenging and 
\begin{figure}[h]
\includegraphics[width=6cm]{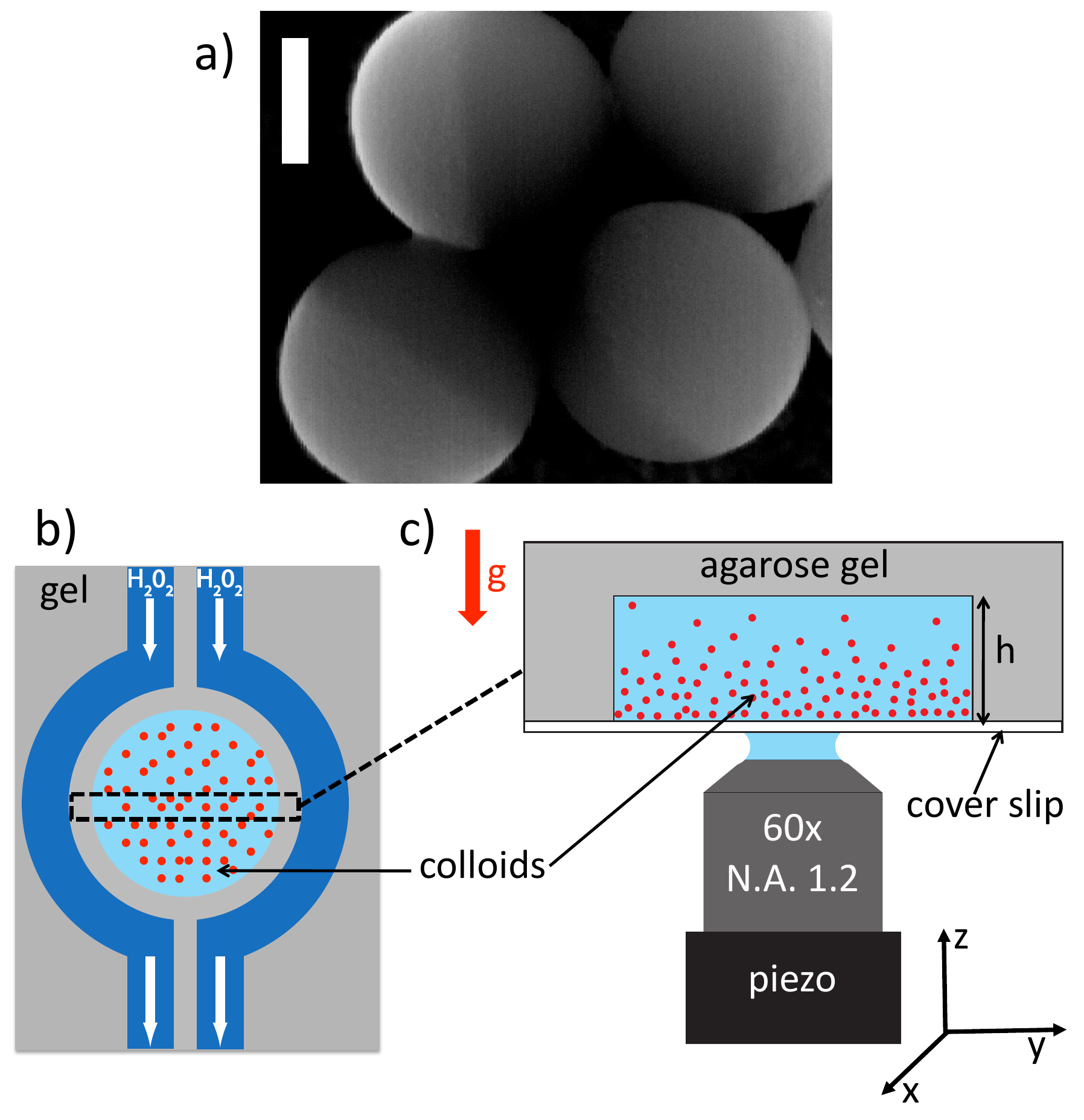}
\caption{(a) SEM picture of janus, platinum-latex, colloids (scale bar 500nm). (b) Sketch of the experimental setup:  a circular microfluidic chamber (diameter $\Phi=650\micron$, height $h\sim80\micron$)  molded in agarose gel (gray areas) contains the $1\micron$ janus colloids. H$_2$O$_2$
feeding is achieved by constant circulation (40$\mu$L/min flow rate) in side channels (dark blue).
(c) The active colloidal suspension is observed through a
piezo-driven high numerical aperture objective (Nikon, Water immersion 60$\times$, NA=1.2) mounted on an inverted microscope.
}
\label{fig-1}
\end{figure}

In this paper we explore experimentally the behavior of a dilute {active suspension} of articial swimmers under an external (gravity) field. This problem was
discussed theoretically in a recent contribution by Tailleur and Cates \cite{Tailleur}. % providing several predictions.
Here the active colloids are powered chemically, following the route proposed by Howse {\it et al.} \cite{Howse}.
The asymetric dismutation of hydrogen peroxide (H$_2$O$_2$) on the colloid itself is used as the driving power, on the basis of a self 
(diffusio-) phoretic motion, induced by the building up of an osmotic pressure gradient at the interface of the colloid \cite{Golestanian,Ajdari,Abecassis}.
Furthermore a specifically designed microfluidic system based on a gel-microdevice technology has been developed, allowing to ensure 
constant renewal of the chemical {\it fuel} (H$_2$O$_2$), as well as removal of {\it waste} products (O$_2$), in an open reactor configuration.
%(mettre open reactor configuration?) in convection free configuration.
%the active suspension is studied in a specifically designed microfluidic system, 
%based on a gel-microdevice technology. 
%This allowed us to bypass the difficulties associated with the 
%draining of the {\it waste} produced by the chemical reaction powering motion (here oxygen).
On the basis of this device, we have explored the behavior of the active suspensions at two complementary levels:
we have first characterized the individual dynamics of active colloids by particle tracking measurements; then we have investigated the behavior of a dilute active suspension of these particles under an external gravity force field, in the same spirit as the historical Jean Perrin experiment \cite{Jean-perrin}. 
This allowed us to connect the micro-dynamics of individual entities to the macroscopic equilibrium behavior of the suspension, allowing to probe the link between fluctuations
and dissipation in this configuration. 
%We show that this  out-of-equilibrium  dilute system is formally ruled by a Schmoluchovsky equation. We relate the {\it individual} dynamics of the motile particles to their {\it global} sedimentation by complementary measurements and show that those microswimmers in a dilute regime behave as ``hot colloids''. 

{\it Active colloids and experimental setup --}
%Self propelled microswimmers are janus particles Platinum-polystirene. The platinum cap catalyses the dismutation of hydrogen peroxide and induces motion thanks to the diffusiophoresis phenomenon
%The janus platinum-PS microswimmers are synthetized  at high enough  concentration of colloids to study the statistical behavior of an assembly. 
A monolayer of commercial fluorescent latex colloids (1 $\mu$m diameter, Molecular Probes F8823) is formed on a silicon wafer by evaporation from $10^3$ dilution in isopropanol (99.99\%, Roth), and coated with 2nm platinum by sputtering, see Fig-\ref{fig-1}-a. Resuspension is achieved by sonication in ultra-pure water (Milli-Q, resistivity 18.2 MOhm.cm$^{-1}$), leading, after centrifugation, to 50 $\mu$L of a janus colloidal solution at $\sim 0.1$\% v/v (about $10^9$ particles/mL).
%
%Active colloids are made starting from 
%a commercial solution of fluorescent latex colloids (F8888, 1 $\micron$, Molecular Probes), diluted 10$^3$  times in isopropanol (99,99\% pure,  Roth). A 
%dilute monolayer of colloids is obtained by evaporating a drop of colloidal solution,
%and then  
%coated with 2nm of platinum by sputering. Fig. 1-a shows a SEM image of the obtained janus colloids.
%The half-platinum coated spheres are then detached from the wafers by  sonication in  distilled water (resistivity 18.2 M$\Omega$.cm$^{-1}$). After centrifugation, we obtain 50~$\mu$L of a janus colloid solution, with a %of concentration around~$5.10^{-3}\%$ are obtained.
% volume fraction $\sim 0.1$\%  (corresponding to $\sim 10^9$ janus colloids per ml). 
 %
 Such janus particles were shown to self-propel in a solution
 of hydrogen peroxide, due to the dismutation of this chemical on the platinum covering half of the colloids \cite{Howse}. 
 %in the range {\color{green} 0.05--0.4 \%}, depending on the sedimentation condition.
  %TOTAL # of colloids in the chamber is from 5 10^5 to 3 10^6: 10^9 colloids/ml
% N(0) =2000 - 20000 
%phi=Vcoll*N(0)/nb_image/Vchamp_camera=0.5µm^3*10^4/100/(150*120*4)~0.05%
%  and stored at $6^{\circ}$C for subsequent use. 
The properties of this colloidal suspension are then investigated in a dedicated microfluidic device, sketched in Fig.1-b-c. It is made of
 a circular microfluidic  chamber (diameter $\Phi=650\micron$, height$\simeq80\micron$, volume $V\simeq30n$L)  molded in agarose gel  \cite{3_channel_Wu,Palacci}, and surrounded by two side channels separated by 125$\mu$m gel walls. 
%A 1.5~$\mu$L drop of the janus colloid solution is deposited on a glass coverslip (thickness 170$\micron$, Roth) and covered with the gel microdevice. 
The central chamber is initially filled with janus colloids, while H$_2$O$_2$ solution at concentration $C_0$ is continuously circulated in lateral channels.
%A flowing solution of hydrogen peroxide with a prescribed concentration $C_0$  is continuously injected in the lateral channels.
This gel microsystem ensures a constant renewal of  H$_2$O$_2$ fuel and removal of  chemical waste products (O$_2$) by diffusion through 
hydrogel walls from the infinite reservoir and sink constitued by the circulating lateral channel. This provides a convection-free environnement 
in the colloids chamber with stable chemical conditions over hours, allowing
%that: {\it(i)} hydrogen peroxide diffuses freely in the gel,  {\it(ii)} the flowing H$_2$O$_2$ solution in the lateral channels acts as an infinite fuel reservoir for the microswimmers, {\it(iii)} the colloids in the chamber are isolated from the outside flow by the screening of the gel, and  {\it(iv)} the oxygen generated by the dismutation of H$_2$O$_2$ is removed by the flow in the lateral channels. 
%Altogether this system allows 
to study %semi-dilute
solutions of active colloids %, with stationary conditions over hours 
(while a similar study would be precluded in a capillary due to the production of oxygen bubbles).
%by the dismutation of H$_2$O$_2$ powering colloidal motion.
 %  is usually very hard to accomplish in capillaries due to gas bubbles production.
In this configuration we then performed measurements of both the dynamics of the colloids and their sedimentation properties, for various concentrations $C_0$ of the fuel.
%, here hydrogen peroxide. %This ensures to suppress any  statistical bias  in the colloid assembly as long as the experiment is not interrupted.
%Observations are carried out on  an inverted microscope (DMI4000B, Leica\textregistered) with a  60$\times$ objective (Nikon, N.A. 1.2, water immersion). 

{\it Colloid micro-dynamics --} %: high speed colloid tracking --}
\begin{figure}
\includegraphics[width=9cm]{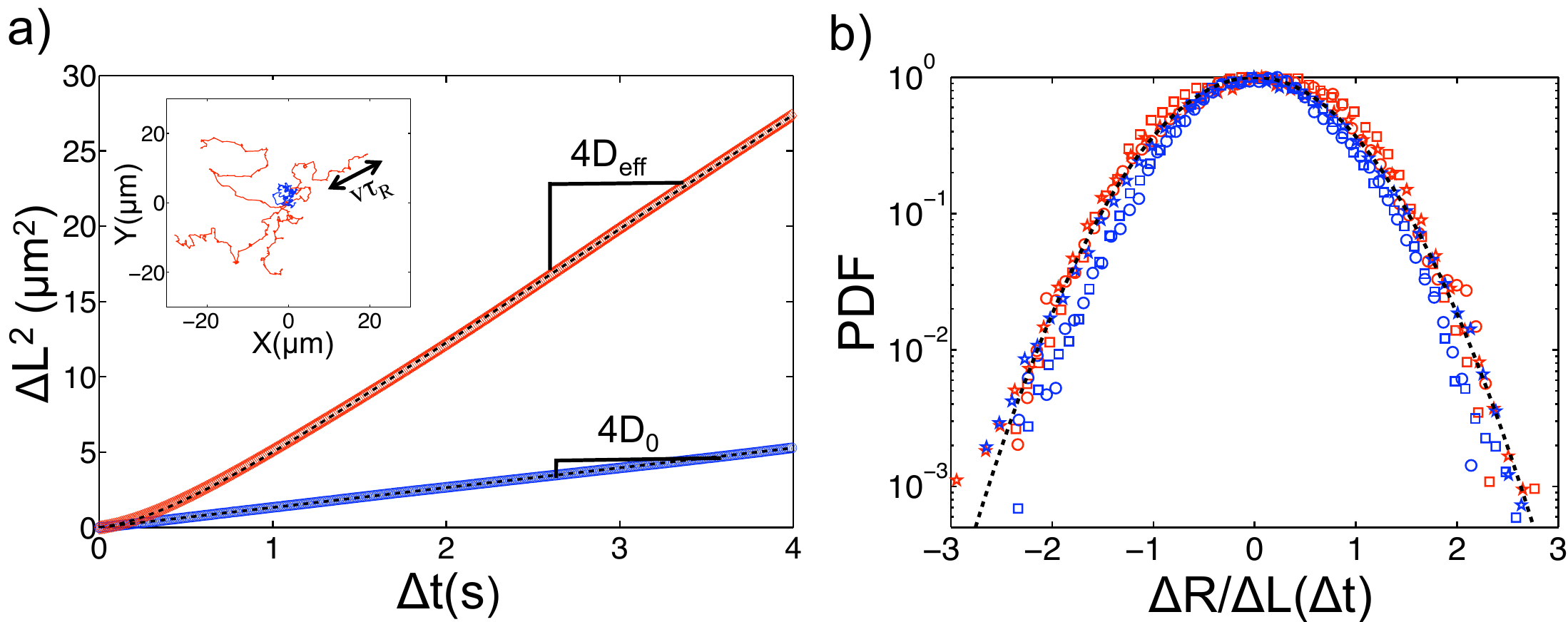}
\caption{(a) Experimental mean squared displacements  $\Delta L^2(\Delta t) $ and 2D trajectories (inset)  for bare (blue) and active colloids (red) in 7.5$\%$ 
H$_2$O$_2$ solution. 
Bare colloids (bottom) show standard diffusion ($\Delta L^2$ linear in time), while the mean squared displacement of active colloids is fitted according to
Eq.(\ref{eq-1}).
The measured diffusion coefficients are  $D_0=0.33\micron^2/s$ for bare and  $D_{\rm eff}=1.9\micron^2/s$ for active colloids.
(b) PDF of particle displacement $\Delta R$ for various lag time $\Delta t$ as a function of the normalized displacement $\Delta R/\Delta L(\Delta t)$. 
$\Delta t$=0.3s($\star$), 1s($\circ$), 3s($\diamond$) for bare (blue) and active colloids (red) in a solution of $7.5\%$ of H$_2$O$_2$. The dashed lines is the gaussian curve.
}
\label{fig-2}
\end{figure}
First, the dynamics of individual active colloids is investigated using 
%Janus colloids dynamics is investigated from 
high speed tracking measurements to resolve temporally the different dynamical regimes (see below). 
%The xenclosed in a microfluidic chamber with  boundary %channels filled by hydrogen peroxide concentration $C_0$. 
For various H$_2$O$_2$ concentrations $C_0$, the two-dimensional (x,y) motion of  colloids is recorded with a high speed camera (Phantom V5) at 100Hz and trajectories are extracted with a single particle tracking algorithm (Spot Tracker, Image J \cite{spottracker}). The mean square displacement of the colloids 
is obtained as $\Delta L^2(\Delta t)=\langle{(\vec{R}(t+\Delta t)-\vec{R}(t))^2}\rangle$ where $\vec{R}(t)$ is the (2D) instantaneous colloid position and the average is performed over time for each individual trajectory and then over an ensemble of trajectories (typically
20).
 %$<\Delta R^2(t)>=<R^2(t)-<R(t)>^2>$ of each  colloid is calculated and averaged in time from the (x,y) coordinates with Matlab and then  averaged over an "ensemble average". 
% The mean-square displacement 
 We plot in Fig.~\ref{fig-2} 
  $\Delta L^2(t) $ as a function of time for bare (non-active) and active colloids in a solution of hydrogen peroxide, confirming
  the impact of injected chemical power on the individual motion of the colloids, in agreement with Ref. \cite{Howse}.
For the bare (non-active) colloids, the dynamics is purely diffusive with a diffusion coefficient $\Delta L^2/ 4\Delta t=D_0=0.34\pm 0.02\micron^2/s$,
which is found to be independent of the H$_2$O$_2$ concentration.
%  We have measured  the mean square displacement for non janus colloids and for different concentrations of $H_2O_2$. We observe a linear  dependency with time lag, as expected for the  diffusive dynamics of colloids at equilibrium, with a diffusion coefficient $\Delta L^2/ 4\Delta t=D_0=0.34\pm 0.02\micron^2/s$. 
Note that the same value and behavior is obtained for janus colloids in the absence of the H$_2$O$_2$ fuel.  
%We have furhermore measured the same value $D_0$ for the diffusion coefficient of janus colloids in pure water, {\it i.e.} {\it without} Êinjected
%  hydrogen peroxide in the lateral channels.
  %This value for the equilibrium diffusion coefficient of 1$\micron$ colloids is close to the equilibrium Stokes Einstein value $D_{SE}=0.44~\micron^2/s$. We infer that the difference is due to an increase of the  viscosity of the hydrogen peroxide solution due to solubilized agarose gel.  This benchmark experiment with janus particles in water allows to check the absence of bias as convective flow.
For  the janus active colloids in a hydrogen peroxide solution, the mean square displacement differs drastically from the equilibrium diffusive dynamics 
and  strongly depends on the fuel concentration. The colloid exhibits ballistic motion at short times,
%At times shorter than the rotational diffusion time $\tau_r\sim0.8s$, colloids exhibit ballistic motion 
$\Delta L^2(t)\sim V^2 t^2$, while at longer times a diffusive regime, $\Delta L^2(t)\sim 4D_{\rm eff} t$, is recovered with an effective diffusion coefficient $D_{\rm eff}$ much larger than the equilibrium coefficient $D_0$. 
As discussed in \cite{Howse,Marchetti}, the active colloids are expected to perform a persistent random walk, due to a competition between ballistic motion under the locomotive power (with a constant swimming velocity $V$), and angular randomization due to thermal rotational Brownian motion. The transition
between the two regimes occurs at the rotational  diffusion time $\tau_r$ of the colloids. %$\tau_r\sim0.8s$
% transition from ballistic to diffusive motion is associated with the 
%{\color{blue} Persistent Random Walk}
%Physically, the colloid swimmer exhibits Brownian diffusion superimposed 
The characteristic ballistic length scale is accordingly $a=V\times \tau_r$. %with V the swimming velocity and $\tau_r$ the rotational diffusion time that represents the decoherence time scale for the swimming direction. 
For time scales long compared to $\tau_r$, the active colloids therefore perform a random random walk with an effective diffusion
 %behavior which combines  thermal Brownian motion and random steps $a$, leading to  an effective diffusion coefficient
  $D_{\rm eff}=D_0 +V^2 \tau_r/4$. %This problem is formally similar to a Langevin equation of a non-Markovian particle with a cutoff time $\tau_r$. 
The full expression of the mean squared displacement at any time is obtained as  \cite{Howse} :
\begin{equation}
\Delta L^2(\Delta t)=4D_0\Delta t +\frac{V^2\tau^2_r}{2}[\frac{2\Delta t}{\tau_r} +e^{\frac{-2\Delta t}{\tau_r}}-1]
\label{eq-1}
\end{equation}
    %We run the same experiment with various concentrations of $H_2O_2$  ($[H_2O_2]=0\%, 3.5\%, 5\%,7.5\%$) with {\it the same sample of colloids in the same microfluidic chamber} while only changing the concentration $C_0$ contained in the syringes. The mean squared displacement of microswimmers with fuel is not linear with time and exhibit more complex behavior (see figure \ref{fig-2}). One can have a basic picture of the asymptotic behavior following the model given by   \cite{Howse:2007fj}. At times much shorter than the rotationnal diffusion $\tau_r\sim 0.8$ (est ce qu on explqique que l'on a rŽpercutŽ l'Žcart de Deq dans $\tau_r$ parce que le produit ne dŽpend que du rayon ?), the colloids present ballistic motion $<\Delta R^2(t)> \propto V^2 t^2$. But at time much longer than $\tau_r$, the propulsion direction is randomized by the thermal noise and one expect to measure an enhanced diffusive behavior $<\Delta R^2(t)> \propto 4*D_{eff} \times t$, with $D_{eff}=D_0+V^2\tau_r/4$.  We extract the diffusion coefficient of the colloids from the equilibrium experiment  in water $D=0.35\micron^2/s$.

We fit  the experimental mean squared displacement  $\Delta L^2(\Delta t)$  using Eq. (\ref{eq-1}) with  the propulsion velocity $V$ as the only free parameter,
 while the value of $D_0$ is taken from the equilibrium diffusion coefficient measured in water and the Stokes expectation is used for $\tau_r$ ($\tau_r= 0.9$s). % (the fit is weakly sensitive on this precise value)
As shown
in Fig. \ref{fig-2}-a, an excellent agreement with the experimental results is found. 
% We want to stress that using this equation we fit $400$ data points with only 2 free parameters with perfect agreement. 
%
%The experimental results are fitted with only 2 free parameters, the propulsion velocity $V$ and the diffusion rotational time $\tau_r$, $D_0$ corresponding to the equuilibrium diffusion coefficient measured in water. We use equation\label{eq-1} to get a fit at any time of the  mean squared displacement  $<\Delta R^2(t)>$ and extract the diffusiophoretic velocity $V$ and the rotationnal time $\tau_r$  of the swimmer for each concentration. We want to stress that we fit $400$ data points with only 2 free parameters with perfect agreement. 
%
 Under the present conditions, the measured propulsion velocities $V$ range from 0.3 $\micron/s$ to 2.5$\micron/s$, corresponding to an increase of the
 effective diffusion coefficient  by a factor up to 5.5 as compared to the bare diffusion.
% {\color{blue}ÊThe rotational diffusion time $\tau_r$ is found to remain within 10\% of its value measured in pure water ($\tau_r=0.9$s).} 
%In addition, we measure that the rotationnal diffusion time $\tau_r$ is an increasing function of $C_0$ and goes from the equilibrium value $0.8s$ at low concentration to $1.2s$ at higher concentration. 
%This concentration dependency of $\tau_r$  interrogates over polarization phenomenon of the janus particle due to the memory of its own solute gradient cloud. Such a Markovian effect differs from the anomalous diffusion prescribed by \cite{golestanian:188305} hardly measurable experimentaly . 
%These results show that the platinum janus particles exhibit propulsion in an environment containing hydrogen peroxide fuel. 
%At long times, the directed propulsion decorrelates and leads to an enhanced diffusion coefficient  $D_{\rm eff}$ that we can evaluate using the values of the fit. In these experiments, the ratio $D_{\rm eff}/D_0$ goes from 1  to 5.5  which corresponds to a large increase of the equilibrium diffusion coefficient: janus microswimmers behave as ``hot colloids".
%This behavior is formally similar to the  brownian diffusion of particles in a bath at temperature $T_{eff}([H_2O_2])$ much higher than the equilibrium bath temperature $T$. In this experiment, the ratio $D_eff/D_0=T_{eff}/T$ goes from 1 at equilibrium in pure water to 3.7 with the highest concentration of fuel we used (7.5$\%$) which corresponds to a large increase of the equilibrium diffusion coefficient.
In the following we use this measure of $D_{\rm eff}$ as a probe of the colloidal 'activity'. %experiment as a probe of 
Finally we have also measured the probability distribution function (PDF) of the colloid displacement $\Delta R=\vert \vec{R}(\Delta t)-\vec{R}(0))\vert$ for a given time lag $\Delta t$, see Fig.\ref{fig-2}-b, both for bare (blue) and active (red) colloid particles. In both cases, the PDF fits very well to a gaussian with variance $\Delta L^2(\Delta t)$ given in Eq.~(\ref{eq-1}).
Even though departures from a gaussian are expected at short time for the persistent random walk, these are within experimental uncertainty.
%While for the the PDFNon janus particles displacement have a gaussian distribution at all time lags whereas janus particles only recover a gaussian distribution after a few seconds. COMPLETER FORME STAT DIFFERENTE...{\color{blue} Persistent Random Walk = no exact result for PDF but here mainly gaussian}

%In order to probe such a "slightly out of equilibrium" description, we perform independant experiments  {\it on the same sample} to determine the response  to a gravitationnal force field of an statistical assembly of microswimmers.   We therefore measure the sedimentation profiles of a population of janus particles in a bath of $H_2O_2$ at various concentration $C_0$.

%{\it Stationary density profile of microswimmers in a gravitational field : a Jean Perrin experiment --}
\begin{figure}
\includegraphics[width=6cm]{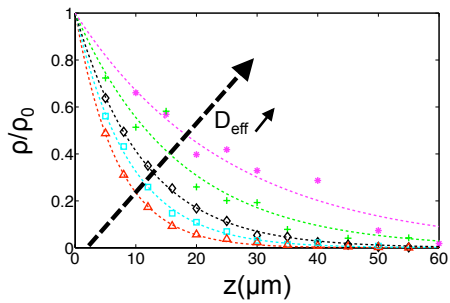}
\caption{Normalized  density profiles $\rho/\rho_0 (z)$  for the active colloidal suspension in stationary state, for increasing swimming activity, {\it i.e.} increasing $D_{\rm eff}$. Experimental data (symbols) are well fitted by an exponential decay $\rho(z)=\rho_0 \exp (-z/\delta_{\rm eff})$, with a sedimentation length $\delta_{\rm eff}$, which strongly depends on (and increases with)
 the swimming activity. Here $\delta_{\rm eff}$ varies from $6\micron$ at equilibrium up to $21\micron$. %with $7.5\%$ of hydrogen peroxide. 
 }
 \label{fig-3}
\end{figure}

{\it Sedimentation of active particles --} 
We now turn to the investigation of the sedimentation of an assembly of such active colloids, probed in a sedimentation experiment, in the 
same spirit as the historical Jean Perrin experiment \cite{Jean-perrin}. 
At thermal equilibrium with a bath at temperature $T$,  a dilute population of colloid with (buoyant) mass $m$ under gravity $g$ exhibits a steady Boltzmann distribution profile $\rho(z)=\rho_0 \exp(-z/\delta_0)$ with $\delta_0= \kt/ mg$ the sedimentation length, which balances gravitational and thermal energy.  
Alternatively this density profile can be seen as the stationary solution of the Smoluchowsky diffusion-convection equation, 
$\partial_tÊ\rho + \nabla\cdot J =0$, with
%\begin{equation}
$J= -D_0 \nabla \rho + \mu\, mg\, \rho$
%\end{equation}
the particle flux, and $D_0$, $\mu$ the colloids diffusion coefficient and mobility.
% the sedimentation length $\delta_0$ is related to the
This leads to the fluctuation-dissipation relationship $D=\kt \mu$. In order to explore the validity of these concepts for the out-of-equilibrium
active suspension, we have measured --
simultaneously to the individual particle tracking experiments -- the density profiles  $\rho(z)$ of the colloids in the microfluidic chamber, for various 
%H$_2$O$_2$ 
fuel concentration $C_0$.
%We therefore use the sedimentation behavior of  self propelled particles to probe statistical physics of such an {\it out of equilibrium system}. 
%As hydrogen peroxide of concentration $C_0$ is continuously injected in the side channels, 
%and we have made sure to be in a stationary state (the sedimentation time without $H_2O_2$ is...). 
Colloid profiles are measured by scanning the chamber using a piezo-mounted microscope objective (PIFOC P-725.2CD, Physik Instrumente).
%microscope, with a 
%%microscope is first focused at $z=0$, {\it i.e.} at the bottom of the microfluidic chamber, then a 
%piezo (P-725 PIFOC, Physik Instrumente) to displace the focus point along the z axis with high accuracy.
At each altitude $z$, a stack of images with lateral dimensions {$150\times 200 \mu$m}  is acquired at 0.3~Hz with a fluorescence camera (Orca, Hamamatsu). On each stack, image analysis is performed with Matlab: first the maximum  fluorescence intensity $I_{\rm foc}$ is determined from  ``in-focus'' colloids, then a $0.3 I_{\rm foc}$ threshold criterion  is applied  in order to discard out-of-focus colloids, thus defining a slice with thickness of {$\pm$ 2$\mu$m} \cite{Born}.
%define the colloids which are contained in the  slice $[z-\delta z; z+\delta z]$, with $\delta z\sim 2\micron$ 
%\footnote{%A good $z$ resolution is obtained  in the focus plane using a high numerical aperture objective, with a corresponding small 
%This $z$ resolution corresponds to a depth of field $\sim{\lambda_0 n}/{{\rm NA}^2} $ with $\lambda_0$ the wavelength and $n$ the refractive index of water.}. 
A stack of 100 images is used to obtain a good statistical convergence.
Overall this allows to obtain the average number of colloids at each altitude $z$.
Note that adsorption of the colloids on the bottom surface
was found to bias the density profiles close to the bottom surface and we thus discarded data from the two {first}Ê slices.
% by bunch of colloids or 
%by surface absorption. %to the surface.
% TOTAL # of colloids in the chamber is from 5 10^5 to 3 10^6: 10^9 colloids/ml
The results of these "Jean-Perrin" experiments are presented in Fig.~\ref{fig-3}. We have checked that  a {\it stationary} state of the sedimentation profile is reached. % (a non-trivial point for an out of equilibrium phenomenon). 
As shown in Fig.~\ref{fig-3}, the density profiles of the active colloidal suspension $\rho(z)$ is decreasing with the altitude $z$ and, as in the thermal case, can be very well fitted by an exponential decay $\rho(z)=\rho_0 \exp (-z/\delta_{\rm eff})$, where $\rho_0$ was used to normalize the different measurements. 
%(with a total number of colloids in the observation window that ranged from $3.10^3$ to $2.10^4$).
%Density profiles are normalized in order to have a unit density at $z=0$, $\rho(0)=1$ \footnote{This is done using the exponential fit to estimate the total number of colloids in the region of interest}. 
%\footnote{Note that the total number of colloids in the observation window ranges typically from $5. \,10^3$ to $3\, 10^4$.} 
%The equilibrium sedimentation length  in pure water is measured as $\delta_0=6\pm2\micron$ in this experiment  \footnote{We want to point that a janus particle is much heavier than the bare  PS colloid (density d=1.05) despite the thinness of the platinum (d=20) cap. This leads to very short sedimentation length. The  obtained value $\delta_0=6µm$ is in  agreement with the expected value of 1$\micron$  PS colloids heavily capped by  $2nm$ platinum.} 
The sedimentation length  $\delta_{\rm eff}$ is found however to depend strongly on the activity of the colloids: $\delta_{\rm eff}$ increases with an increased propulsion of the colloids, {\it i.e.} injected energy, as measured (independently) by their effective diffusion coefficient $D_{\rm eff}$.  %on the propulsion  of the janus particles that is added to the thermal motion to balance the gravitationnal field.
The exponential decay suggests that in the present limit of a dilute active suspension, the active colloids still obey an effective Smoluchowsky 
equation, with the current replaced by $ J= -D_{\rm eff} \nabla \rho + \mu\, mg\, \rho$. This predicts a sedimentation length in the form
\begin{equation}
\delta_{\rm eff}= {1\over v_T} \times D_{\rm eff},
\label{FDT_eff}
\end{equation}
where $v_T=\mu\, mg$ is the sedimentation velocity in the gravity field.
%
% In addition, from the exponential decay of the density profile of colloids $\rho(z)=\rho_0 exp (-z/\delta_{\rm eff})$ we can derive a Schmoluchowsky description and define an effective global diffusion coefficient  $D_{g}$, $\rho(z)=\rho_0 exp (-mgz\mu/D_{g})$ where $\mu$ is the mobility of the colloid. 
\begin{figure}
\includegraphics[width=7cm,height=5cm]{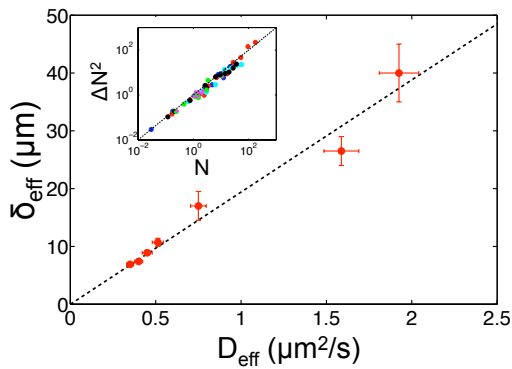}
\caption{ Sedimentation length $\delta_{\rm eff}$ as a function of  effective diffusion $D_{\rm eff}$ extracted from Figs. \ref{fig-2} and \ref{fig-3} respectively.
 %the global sedimentation and the individual tracking experiments. 
  The dashed line is a linear fit, $\delta_{\rm eff}= \alpha \times D_{\rm eff}$ as expected from Eq. (\ref{FDT_eff}). The measured slope $\alpha=${19.5 $\pm 2$} $\mu$m$^{-1}.$s is furthermore 
in good agreement with the expected value $\alpha={1/ v_T} =\delta_0/D_0= {16.2 \pm 2.5}$ $\mu$m$^{-1}$.s, with $\delta_0$ and $D_0$ measured
independently in the absence of injected H$_2$O$_2$.
%Data for the sedimentation length and diffusion coefficient are normalized by their values in water, measured independently.
%Experimental data points are found to collapse on the master curve $y=x$ (dashed line). 
%This shows the validity o global indivuale self-propelled particles with an effective diffusion coefficient $D_{eff}$ increasing with the energy injected by the swimmers transducing chemicals into mechanical energy. 
 {\it Inset:} mean squared number of colloids in each slice $\Delta N^2$ versus the averaged number $N$ for various swimming activity. Symbols are similar to Fig.\ref{fig-3}.
The dashed line is the prediction for equilibrium system $\Delta N^2=\kappa\cdot N$, with $\kappa=1$ for dilute systems.
}
\label{fig-4}
\end{figure}

%We can also make an anology with sedimentating janus particles of the same mass at thermal equilibrium with a bath at temperature $T_{sed}$. This temperature depends on the hydrogen peroxide concentration and increases with the injected energy.
We have checked this relationship by plotting the sedimentation length measured from the colloid density profiles, against the 
effective diffusion coefficient measured in the individual tracking measurements. As shown in Fig. \ref{fig-4}, 
the predicted proportionnality between sedimentation length and effective diffuson coefficient is demonstrated experimentally for all propelling activities, 
%with a slope agreeing with the theoretical value in Eq (2).
%
%the sedimentation length
%is found to be directely proportional to the effective diffusion coefficient whatever the propelling  activity, 
%thereby assessing the validity of above link betwen $\delta_{\rm eff}$ and $D_{\rm eff}$ in Eq. (\ref{FDT_eff}). Furthermore the 
with a proportionality constant which is furthermore found to agree with its expected value in Eq. (\ref{FDT_eff}).
%Eq. (\ref{FDT_eff}). 
This effectively connects the micro-dynamics of the active colloids to their global stationary profile. 
%in a modified form of the fluctuation-dissipation relationship. 

Finally, we have also measured in the previous experiments the mean squared number of particles, $\Delta N^2$(z) in each slice $\delta z$ for various altitudes $z$. We plot in Fig.~\ref{fig-4}{-(inset)} $\Delta N^2$(z) versus the averaged number $N(z)$ for all fuel concentrations. Data are found to collapse onto a single curve $\Delta N^2=\kappa\cdot N$ with $\kappa=1$.
This shows that under the present dilute conditions, the active suspension behaves like an equilibrium (ideal) solution, and do not exhibit anomalously large
fluctuations ($\Delta N^2 \sim N^\gamma$, with $\gamma >1$), as previously predicted \cite{ramaswamy}, however for denser active systems.
% from which we can define an effective temperature $T_{eff}$: $D_{eff}=\mu \times k_{\text{B}}T_{eff}$ where $\mu$ is the mobility of the colloid.
% with the effective temperature increasing with the injected energy. 
%The stationary sedimentation measurements show that the profiles of an assembly of self propelled particles are well described by an exponential decay $\rho(z)=\rho_0 exp (-z/\delta_{\rm eff})$ from which we can extract an effective diffusion coefficient. 

%{\it Discussion--}
Our results show that in the present regime, the active colloids behave merely as ``hot'' colloids, with an effective temperature $k_B T_{\rm eff}=D_{\rm eff}/\mu$ much larger than the bare temperature, and do obey an effective Smoluchowsky equation. In the '{\it run and tumble}' model of particles under external fields discussed by Tailleur and Cates, this behavior is expected in the regime
where the swim speed $V$ is larger than the sedimentation velocity $v_T$. Here $V \sim \mu$m/s, while $v_T =1/\alpha\sim 50.10^{-3} \mu$m/s -- see Fig.~\ref{fig-4}-- , and one has $V/v_T\gg 1$ in the present experiments.
% typical propulsive 'force' $f_p$ acting on the colloids is quite larger than gravity $mg$ \cite{Tailleur}. In our case, $f_p$ can be estimated from the
%viscous stresses within the interface diffuse layer at the colloid's surface where the propulsive power takes place, with nanometric thickness  $\lambda$: 
%$f_p \sim \eta V/\lambda \times 4 \pi R^2$ \cite{Ajdari} . We find here $f_p/mg \sim 10^2$, 
Our experimental results thus validate the theoretical expectations in this regime.

%mettre l'idŽe que artificial donc mechanisms under control + first "collective" au sens population mme si on peut les traiter comme indŽpendant ici et que c'est un first step vers "true" collective behvior. Une partie de ces idŽes sont dans le texte actuel mais j'ai l'impression que ce fil qui fait le point fort n'est pas totalement repris ici
To conclude, we have proposed efficient experimental tools allowing to explore the properties of {\it active colloidal suspensions} under controlled
and tunable conditions. 
This involves artificial chemically active colloids, studied in a dedicated gel microfluidic system.
%microfluidic technology.
%constitutes a great experimental tool to investigate the out-of-equilibrium behavior of active supensions.
%The possibility of investigating active colloidal suspension opens up 
%Various extension
This opens up many perspectives towards a thorough exploration of the out-of-equilibrium behavior of active supensions, in particular for higher volume fractions, where exotic behavior was predicted theoretically, but not observed up to now with artificial active fluids. This involves collective effects, the emergence of flow order, anomalous fluctuations \cite{ramaswamy}, nematic ordering \cite{Marchetti}, etc. But beyond these predicted behaviors, it is interesting to note that the present colloids not only interact via hydrodynamic
flows, but also via {\it chemical} interactions, through the spatial (diffusive) extension of the consummed fuel. This is expected to affect the 
many-body behavior of the suspension, with couplings which have not been considered up to now in the litterature. Work along these lines is in progress.
%
%A key feature of the present 
%{\color{blue} Now we have demonstrated possibity to study active suspensions, we will now go to manybody effects and denser suspensions. 
%Nematic ordering is expected. Work in progress}

%{\it Acknowledgement}
We thank E. Vigier, H. F\'eret, A. Piednoir, J. Gr\'egoire, J.-M. Beno\^it. 
We acknowledge %assistance from Physik Instrumente and 
support from R\'egion Rh\^one-Alpes under program CIBLE.

%\bibliographystyle{unsrt}
%\bibliography{./Biblio_Jean_Perrin.bib}

%\bibliography{./biblio}

\begin{thebibliography}{99}

\bibitem{ramaswamy} R.A. Simha, S. Ramaswamy, {\it Phys. Rev. Lett.} {\bf 89},Ê 058101 (2002).
%S. Ramaswamy and M. Rao, {\it New J. Phys.} {\bf 9} 423 (2007).

\bibitem{Vicsek}
T.~Vicsek, A.~Czir{\'o}k, E.~Ben-Jacob, I.~Cohen, and O.~Shochet.
%\newblock Novel type of phase transition in a system of self-driven particles.
\newblock {\em Phys. Rev. Lett.} {\bf 75}, 1226 (1995).

\bibitem{Joanny}
K. Kruse, J. F. Joanny, F. J\"ulicher, J. Prost, and K. Sekimoto, {\it Eur. Phys. J. E} {\bf 16}, 5 (2005).

\bibitem{ignacio}
I.~Llopis and I.~Pagonabarraga.
%\newblock Dynamic regimes of hydrodynamically coupled self-propelling particles.
\newblock {\em Europhys. Lett.} {\bf 75}, 999 (2006).


\bibitem{Chate}
%G. Gr\'egoire and H. Chat\'e, {\it Phys. Rev. Lett.} {\bf 92}Ê 025702 (2004).
H. Chat\'e, F. Ginelli and R. Montagne, {\it Phys. Rev. Lett.} {\bf 96}Ê 180602  (2006).

\bibitem{diLeonardo}ÊL. Angelani, R. Di Leonardo, and G. Ruocco,
{\it Phys. Rev. Lett.} {\bf 102}Ê 048104 (2009).

\bibitem{Marchetti} A. Baskaran and M.C. Marchetti, {\it Phys. Rev. Lett.} {\bf 101}, 268101 (2008)

\bibitem{Tailleur}
J.~Tailleur and M.E.~Cates 
%\newblock Sedimentation, trapping, and rectification of dilute bacteria.
\newblock {\em Europhys. Lett.}  {\bf 86}, 60002, 2009.


\bibitem{Lauga}
E. Lauga and T.R. Powers,
{\it Rep. Prog. Phys.} {\bf 72} 096601 (2009)

\bibitem{Lauga2}
A.P. Berke, L. Turner, H.C. Berg, E. Lauga, {\it Phys. Rev. Lett.} {\bf 101}Ê 038102 (2008).

\bibitem{leptos}
K.~C. Leptos, J.~S. Guasto, J.~P. Gollub, A.~I. Pesci, and R.~E. Goldstein.
%\newblock Dynamics of enhanced tracer diffusion in suspensions of swimming
%  eukaryotic microorganisms.
\newblock {\em Phys. Rev. Lett.} {\bf 103} 198103 (2009).

\bibitem{Libchaber2000}
X.-L. Wu and A.~Libchaber.
%\newblock Particle diffusion in a quasi-two-dimensional bacterial bath.
\newblock {\em Phys. Rev. Lett.} {\bf 84}, 3017 (2000).

\bibitem{Cisneros}
L.~Cisneros, R.~Cortez, C.~Dombrowski, R.~Goldstein, and J.~Kessler.
%\newblock Fluid dynamics of self-propelled microorganisms, from individuals to
%  concentrated populations.
\newblock {\em Exp. Fluids}, {\bf 43} 737, 2007.

\bibitem{Rafai} S. Rafa\"\i, L. Jibuti, P. Peyla, {\it Phys. Rev. Lett.} {\bf 104}, 098102 (2010)


\bibitem{Howse}
J.~R. Howse, {\it et al.}
% R~A~L Jones, A~J Ryan, T~Gough, R~Vafabakhsh, and R~Golestanian.
%\newblock Self-motile colloidal particles: From directed propulsion to random
%  walk.
\newblock {\em Phys. Rev. Lett.} {\bf 99}, 048102 (2007).

\bibitem{Bibette} R. Dreyfus, {\it et al.}
%J. Bibette, 
{\it Nature} {\bf 437} 862 (2005)

\bibitem{Paxton}ÊW.F. Paxton {\it et al.} {\it J. Am. Chem. Soc.}Ê{\bf 26}, 13424 (2004)

\bibitem{Golestanian} R. Golestanian, T.B.  Liverpool, A. Ajdari, 
{\it Phys. Rev. Lett.} {\bf 94}, 220801 (2005)
%Propulsion of a Molecular Machine by Asymmetric Distribution of Reaction Products

\bibitem{Ajdari} A. Ajdari and L. Bocquet, {\it Phys. Rev. Lett.} {\bf 96}, 186102 (2006).

\bibitem{Abecassis} B. Ab\'ecassis, C. Cottin-Bizonne, C. Ybert, A. Ajdari, L. Bocquet, 
{\it Nature Mat.} {\bf 7} 785 (2008).


\bibitem{Jean-perrin}
J.~Perrin.
%\newblock Mouvement brownien et r{\'e}alit{\'e} mol{\'e}culaire.
%\newblock {\em Annales de Chimie et de Physique}, 
\newblock {\em Ann. Chim. Phys.} (Paris), 
{\bf 8}, 1 (1909).

\bibitem{3_channel_Wu}
J.~Diao, {\it et al.}
%L.~Young, S.~Kim, E.~A Fogarty, S.~M Heilman, P.~Zhou, M.~L .Shuler,
%  M.~Wu, and M.~P. DeLisa.
%\newblock A three-channel microfluidic device for generating static linear
%  gradients and its application to the quantitative analysis of bacterial
%  chemotaxis.
\newblock {\em Lab on a Chip}, {\bf 6}, 381 (2006).

\bibitem{Palacci} J. Palacci, B. Ab\'ecassis, C. Cottin-Bizonne, C. Ybert, L. Bocquet, 
{\it Phys. Rev. Lett.} {\bf 104} 138302 (2010).

\bibitem{spottracker}
F.~Hediger S.M. Gasser M.~Unser D.~Sage, F.R.~Neumann,
%\newblock Automatic tracking of individual fluorescence particles: Application
%  to the study of chromosome dynamics.
\newblock {\em IEEE Transactions on Image Processing}, {\bf 14}, 1372, (2005).

\bibitem{Born} M. Born and E. Wolf, {\it Principle of optics} (Cambridge University Press, 7$^{\rm th}$ Ed., 1999)

\end{thebibliography}

\end{document}